\documentclass[
    twocolumn,
    floatfix
]{aastex631}

\usepackage{mathtools}
\usepackage{amsmath}
\usepackage{savesym}
\savesymbol{tablenum}
\usepackage{siunitx}
\restoresymbol{SIX}{tablenum}

\usepackage{booktabs}

\usepackage{cleveref}

\defcitealias{chen_constraining_2019}{C19}
\defcitealias{middleton_massive_2021}{M21}
\defcitealias{agazie_nanograv_2023}{NG15}
\defcitealias{graham_systematic_2015}{G15}
\defcitealias{shen_bolometric_2020}{S20}

\DeclareSIUnit\Msun{M_{\odot}}

\graphicspath{{./}}

\turnoffedit

\begin{document}

\title[Quasars can Signpost SMBHBs]{Quasars can Signpost Supermassive Black Hole Binaries}

\author[0000-0002-5557-4007]{J. Andrew Casey-Clyde}
\affil{Department of Physics, University of Connecticut, 196 Auditorium Road, U-3046, Storrs, CT 06269-3046, USA}
\affil{Department of Physics, Yale University, 217 Prospect Street, New Haven, 06511, CT, USA}
\correspondingauthor{Andrew Casey-Clyde}
\email{andrew.casey-clyde@uconn.edu}

\author[0000-0002-4307-1322]{Chiara M. F. Mingarelli}
\affil{Department of Physics, Yale University, 217 Prospect Street, New Haven, 06511, CT, USA}

\author[0000-0002-5612-3427]{Jenny E. Greene}
\affil{Department of Astrophysical Sciences, Princeton University, Peyton Hall, 4 Ivy Lane, Princeton, NJ 08544, USA}

\author[0000-0003-4700-663X]{Andy D. Goulding}
\affil{Department of Astrophysical Sciences, Princeton University, Peyton Hall, 4 Ivy Lane, Princeton, NJ 08544, USA}

\author[0000-0002-3118-5963]{Siyuan Chen}
\affil{Shanghai Astronomical Observatory, Chinese Academy of Sciences, 80 Nandan Road, Shanghai, 200030, PR China}
\affil{Kavli Institute for Astronomy and Astrophysics, Peking University, 5 Yiheyuan Road, Beijing, 100871, PR China}

\author[0000-0002-1410-0470]{Jonathan R. Trump}
\affil{Department of Physics, University of Connecticut, 196 Auditorium Road, U-3046, Storrs, CT 06269-3046, USA}

\begin{abstract}
Supermassive black holes (SMBHs) are found in the centers of massive galaxies, and galaxy mergers should eventually lead to SMBH mergers. Quasar activity has long been associated with galaxy mergers, so here we investigate if supermassive black hole binaries (SMBHBs) are preferentially found in quasars. Our multimessenger investigation folds together a gravitational wave background signal from NANOGrav, a sample of periodic AGN candidates from the Catalina Real-Time Transient Survey, and a quasar mass function, to estimate an upper limit on the fraction of quasars which could host a SMBHB. We find at 95\% confidence that quasars are at most five times as likely to host a SMBHB as a random galaxy. Pulsar timing arrays may therefore be more likely to find SMBHBs by prioritizing quasars over a random selection of galaxies in their searches.
\end{abstract}


\section{Introduction} \label{sec:intro}
The primary goal of pulsar timing array (PTA) experiments is to detect low-frequency gravitational-waves (GWs).
All current PTA experiments have now found evidence for a GW background (GWB) in their pulsar data (\citealp{agazie_nanograv_2023}, NG15 hereafter; \citealp{reardon_search_2023,antoniadis_second_2023,xu_searching_2023}).
We expect the primary source of the GWB to be the cosmic population of inspiralling supermassive black hole binaries (SMBHBs).
In this low-frequency GW regime, the binaries' evolution is so slow that at any given moment there could be hundreds of thousands of SMBHBs emitting GWs, thus creating a GWB.
Though most SMBHBs are expected to be unresolvable, a handful of particularly loud and nearby systems could  be detectable as continuous gravitational wave (CW) sources \citep{mingarelli_local_2017, xin_multimessenger_2021}.

Galaxy mergers have long been associated with quasar activity \citep{sanders_ultraluminous_1988,volonteri_assembly_2003,granato_physical_2004,hopkins_cosmological_2008} -- it therefore follows that some quasars may host SMBHBs.
Targeted GW searches, where the host galaxy is known, offer up to an order of magnitude improvement on the strain upper limits over all-sky GW searches~\citep{arzoumanian_multimessenger_2020}. It is therefore crucial to understand which galaxies are most likely to host SMBHB systems, in the new hunt to detect CWs from individual SMBHB systems.

Given the link between galaxy mergers and quasar activity, in this paper we investigate if quasars can signpost SMBHB systems.
We start from a catalog of quasars with periodic light curves, from the Catalina Real-time Transient Survey \citep[CRTS,][G15 hereafter]{graham_systematic_2015}, which hydrodynamics simulations suggest may be the signature of a SMBHB \citep{dorazio_accretion_2013,farris_binary_2014,shi_threedimensional_2015}.
Since intrinsic quasar variability can mimic periodicity \citep[see, e.g.][]{vaughan_false_2016,witt_quasars_2022,davis_reliable_2024} we use NANOGrav's GWB measurement \citep{agazie_nanograv_2023} to constrain the maximum number of genuine binaries in the CRTS catalog.
We then compare the GWB-constrained CRTS catalog to the quasar population. This, in turn, enables us to constrain the fraction of quasars that can host a SMBHB.
Finally, to understand if quasars can preferentially host SMBHBs we compare the fraction of quasars hosting a SMBHB to the fraction of all galaxies hosting a SMBHB. Both these quantities are constrained by NANOGrav's GWB measurement.

This paper is organized as follows:
In Section \ref{sec:crts} we describe the CRTS sample.
In Section \ref{sec:mfs} we review the mass functions used and describe how we calculate binary occupation fractions for galaxies in general and for quasars.
In Section \ref{sec:results} we present our results, including the expected number of genuine binaries in CRTS, and whether quasars are more likely to host SMBHBs than random galaxies.
In Section \ref{sec:conclusion} we summarize our main findings and provide directions for future research.

Throughout this paper we use natural units where $G = c = 1$.
We assume a standard Lambda CDM cosmology with Hubble parameter $h_{0} = 0.7$, constant $H_{0} = 70~\rm{km}~\rm{s}^{-1}~\rm{Mpc}^{-1}$, and energy density ratios $\Omega_{M} = 0.3$, $\Omega_{k} = 0$, and $\Omega_{\Lambda} = 0.7$.

\section{CRTS Sample}
\label{sec:crts}

    CRTS is a time-domain optical survey covering $\sim 33,000 \; \mathrm{deg}^{2}$ to a depth of $V \sim 19-21.5$ \citepalias{graham_systematic_2015}.
    It has produced light curves for millions of objects, including AGN.
    Simulations suggest periodic AGN light curves may trace binary activity \citep[e.g.][]{farris_binary_2014}
    This is due to periodic accretion from the circumbinary disk onto the binary \citep{dorazio_accretion_2013}, overdense lumps in the circumbinary disk \citep{farris_binary_2014}, and Doppler boosting of light from the minidisks around each SMBH \citep{dorazio_relativistic_2015}.
    Time-domain surveys such as CRTS are therefore crucial in searches for electromagnetic counterparts to SMBHBs.
    
    In \citetalias{graham_systematic_2015} the authors searched for periodic quasar light curves in CRTS, which had a nine year baseline at the time.
    \citetalias{graham_systematic_2015} identified $334,446$ spectroscopically confirmed quasars in CRTS by cross-matching with SDSS Data Release 12 \citep{paris_sloan_2017} and the Million Quasars catalog\footnote{http://quasars.org/milliquas.htm}.
    $243,486$ of these quasars had light curves with sufficient quality to search for periodicity.
    \citetalias{graham_systematic_2015} identified $111$ periodically varying quasars in this sample, proposing them as SMBHB candidates.

    Starting from these candidates, we constrain the population of binary quasars, i.e., SMBHBs with associated quasar activity from a circumbinary accretion disk.
    We first assess the completeness of the CRTS sample, including the PTA frequencies accessible to CRTS via periodic quasar light curves.
    Since we expect $z \approx 1.5$ encompasses $95 \%$ of the GWB \citep{sesana_systematic_2013}, we initially consider the 95 candidates within this volume.

\subsection{CRTS Completeness}
\label{sec:complete}

Here we follow the arguments from \citet{sesana_testing_2018} to determine the effective completeness of CRTS.
Whereas \citet{sesana_testing_2018} consider the full CRTS sample, our calculation is limited to the candidates within the GWB volume $z \leq 1.5$.

CRTS is limited by sky coverage, data quality, and survey depth (\citealp{drake_probing_2013}, \citetalias{graham_systematic_2015}, \citealp{sesana_testing_2018}).
While CRTS covers $\sim 80 \%$ of the sky, only $67/95$ of the CRTS candidates we consider are spectroscopically confirmed as quasars by SDSS, which has $\sim 23 \%$ sky coverage.
Assuming complete identification in SDSS, we expect $67 / 0.23 \approx 291$ candidates within $z \leq 1.5$ in the whole sky.
This suggests the effective sky coverage of CRTS is $95 / 291 \approx 33 \%$.
For comparison, \citet{sesana_testing_2018} estimate the effective sky coverage of CRTS as $35.5 \%$, assuming $25 \%$ sky coverage in SDSS.

\citetalias{graham_systematic_2015} additionally reject $\sim 25 \%$ of quasars for poor temporal coverage in their light curves, further limiting completeness to $\sim 24 \%$ \citep[$\sim 26 \%$ in][]{sesana_testing_2018}.
Finally, CRTS sees to a depth of $V \sim 19 - 21.5$ \citepalias{graham_systematic_2015}.
We exclude $7$ CRTS candidates which fall below the $V = 19$ flux completeness limit, leaving us with $88$ flux complete candidates within $z \leq 1.5$.

\subsection{PTA Band Accessibility}
\label{sec:pta}

\begin{figure*}
        \centering
        \includegraphics[width=0.7\linewidth]{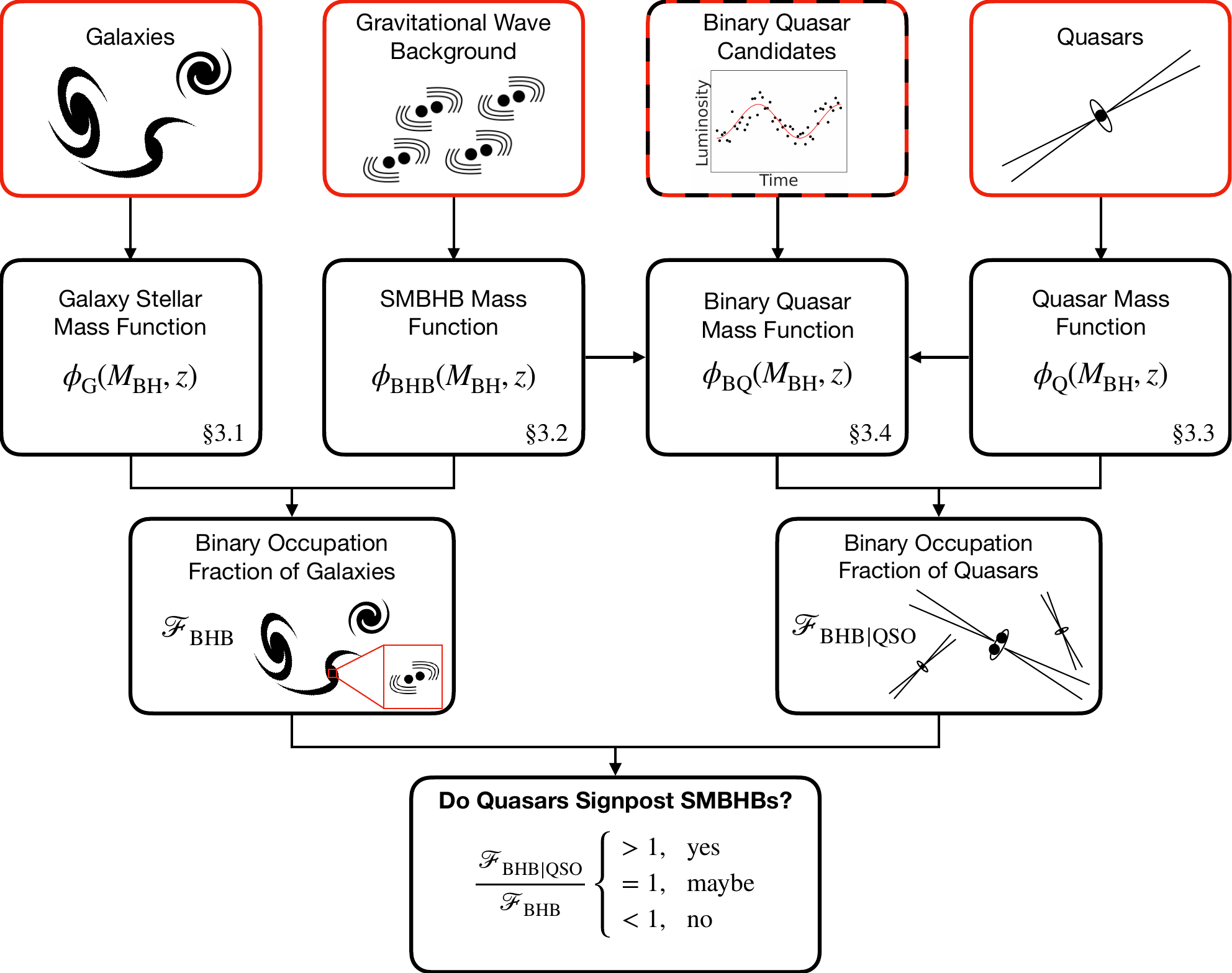}
        \caption{Here we outline how to determine if quasars are likely to signpost binaries. Observationally constrained quantities, such as galaxies, quasars, and the GWB, have red boxes, while theoretical quantities are black.
        Binary quasars have a red and black box because while the periodic quasar light curves are observed, theory motivates their status as SMBHB candidates.
        We start with a galaxy stellar mass function, i.e., the number of galaxies per unit volume and stellar mass (Section \ref{sec:bhmf}).
        We then use the measurement of the GWB amplitude to constrain the SMBHB mass function (Section \ref{sec:bhbmf}).
        From this we derive a SMBHB occupation fraction by comparing to the galaxy stellar mass function (Section \ref{sec:fractions}).
        Similarly, to calculate the fraction of quasars that host SMBHBs we use a quasar mass function (Section \ref{sec:qmf}) and a binary quasar mass function (Section \ref{sec:bqmf}). 
        Binary quasar candidates, identified via electromagnetic observations, will eventually be confirmed or rejected by PTA observations.
        }
        \label{fig:duty_cycle_flowchart}
\end{figure*}

CRTS only considers candidates with at least $1.5$ periods in the nine year baseline \citepalias{graham_systematic_2015}.
This corresponds to a maximum orbital period $P \lesssim 6 \; \rm{yr}$, or $f_{\rm GW} \gtrsim 10^{-8} \; \mathrm{Hz}$ \citep{peters_gravitational_1963}, assuming the quasar light curve period traces the binary orbital period\footnote{Hydrodynamic studies suggest this is true for unequal mass binaries, though the maximum mass ratio this holds for depends on the simulation details \citep[e.g.,][]{dorazio_accretion_2013,farris_binary_2014,miranda_viscous_2017,munoz_circumbinary_2020}.}, similar to previous studies (\citetalias{graham_systematic_2015}, \citealp{sesana_testing_2018,xin_multimessenger_2021}).
We exclude one candidate, SDSS J113916.47+254412.6, which we find is below this lower limit, leaving a total of $87$ binary quasar candidates in our analysis.
Since the differential frequency distribution of circular binaries scales as $f_{\rm GW}^{-11 / 3}$ \citep{peters_gravitational_1964}, the total number of binaries integrated over some frequency interval scales as $f_{\rm GW, min}^{-8/3}$, where $f_{\rm GW, min}$ is the lowest frequency considered.
This implies that CRTS is sensitive to $(10^{-8} \; \mathrm{Hz})^{-8/3} / (10^{-9} \; \mathrm{Hz})^{-8/3} \approx 0.2 \%$ of the PTA band, $10^{-9} \leq f_{\rm GW} \leq 10^{-7} \; \mathrm{Hz}$.

\section{Mass Functions}
\label{sec:mfs}

To compute binary occupation fractions for galaxies in general, and quasars in particular, we must model the SMBH, SMBHB, quasar, and binary quasar populations, \autoref{fig:duty_cycle_flowchart}.
We model these populations with their respective mass functions, $\phi = d \Phi(M, z) / d \log M$, where $\Phi$ is the comoving number density of SMBHs, SMBHBs, quasars, or binary quasars, $M$ is the total mass of each, and $z$ is redshift.
Here we detail the construction of these mass functions, which we use to predict what fraction of quasars could host SMBHBs.
Since the minimum SMBH mass contributing to the GWB is $10^8 \; M_\odot$ \citep{casey-clyde_quasarbased_2022}, the mass functions we consider have $10^{8} \leq M_{\rm BH} \leq 10^{10.5}~\rm{M}_{\odot}$.
Uncertainties in all of our results are computed via $1,000$ Monte Carlo realizations of all models.

\subsection{Supermassive Black Hole Mass Function}
\label{sec:bhmf}

We construct the SMBH mass function as in \citet{sesana_stochastic_2008}, \citet{sesana_systematic_2013}, and \citet[C19 hereafter]{chen_constraining_2019}.
Following \citetalias{chen_constraining_2019}, we start from a galaxy stellar mass function, $\phi_{*} = d\Phi_{*} / d \log M_{*}$, modeled as a redshift evolving Schechter function:
    \begin{equation}
        \label{eq:gsmf}
        \phi_{*}(M_{*}, z) = \ln10 \Phi_{*}(z) \left(\frac{M_{*}}{M_{0}}\right)^{1 + \alpha_{*}(z)} \exp\left(-\frac{M_{*}}{M_{0}}\right) \, .
    \end{equation}
Here $M_{*}$ is the galaxy stellar mass while $\log \Phi_{*}(z) = \phi_{0} + \phi_{1} z$ and $\alpha_{*}(z) = \alpha_{0} + \alpha_{1} z$ are phenomenological functions of $z$.
    
We next calculate the SMBH mass function from the galaxy stellar mass function in two steps.
First we assume scaling relations between bulge mass, $M_{\mathrm{bulge}}$, and $M_{*}$, accounting for differences in the bulge fractions of early and late type galaxies.
Then we adopt a scaling between $M_{\mathrm{bulge}}$ and SMBH mass, $M_{\mathrm{BH}}$.

To account for differences in the $M_{\mathrm{bulge}} - M_{*}$ scaling for early and late type galaxies, we adopt different scaling relations for these populations.
For early-type galaxies we assume a phenomenological scaling from \citetalias{chen_constraining_2019} \citep[cf.][]{bernardi_systematic_2014,sesana_selection_2016}:
\begin{widetext}
\begin{equation}
\label{eq:mg-mbulge}
    \frac{M_{\mathrm{bulge}}}{M_{*}} = \begin{cases}
        \frac{\sqrt{6.9}}{\left[\log \left(M_{*} / \mathrm{M}_{\odot}\right) - 10\right]^{1.5}} \exp\left[\frac{-3.45}{\log \left(M_{*} / \mathrm{M}_{\odot}\right) - 10}\right] + 0.615 & \text{if } \log\left( M_{*} / \mathrm{M}_{\odot}\right) > 10 \\
        0.615 & \text{if } \log \left(M_{*} / \mathrm{M}_{\odot}\right) \leq 10 \, ,
    \end{cases}
\end{equation}
\end{widetext}
with an intrinsic dispersion of $0.2 \; \mathrm{dex}$ \citep{sesana_selection_2016}.
The probability, $P_{\mathrm{ET}}(M_{\mathrm{bulge}} \vert M_{*})$, that an early-type galaxy with stellar mass $M_{*}$ has a bulge of mass $M_{\mathrm{bulge}}$ is thus log-normally distributed with log-space mean given by the logarithm of \autoref{eq:mg-mbulge} and dispersion $0.2$.
For late-type galaxies we assume $P_{\mathrm{LT}}(M_{\mathrm{bulge}} \vert M_{*}) = \mathcal{F}_{\mathrm{bulge, LT}} / M_{*}$, where $\mathcal{F}_{\mathrm{bulge, LT}}$ is the bulge fraction of a late type galaxy which we assume is uniformly distributed between $0.1$ and $0.3$ \citep{sesana_systematic_2013}.

We can then express $\phi_{*}$ in terms of bulge mass by convolving the galaxy stellar mass function with these probabilities as
% \begin{widetext}
\begin{equation}
\label{eq:bulgemr}
\begin{split}
    \phi_{\mathrm{bulge}} &= \int \left[ \mathcal{F}_{\mathrm{ET}} P_{\mathrm{ET}}(M_{\mathrm{bulge}} \vert M_{*})\right. \\
    & \qquad \left. + (1 - \mathcal{F}_{\mathrm{ET}}) P_{\mathrm{LT}}(M_{\mathrm{bulge}} \vert M_{*}) \right] \phi_{*} d \log M_{*} \, ,
    \end{split}
\end{equation}
% \end{widetext}
where $\mathcal{F}_{\mathrm{ET}}$ is the fraction of galaxies which are early type galaxies.
We adopt the $M_{*}$- and $z$-dependent $\mathcal{F}_{\mathrm{ET}}$ for massive galaxies from \citet{huertas-company_galaxy_2024}, which used the James Webb Space Telescope (JWST) to constrain galaxy morphology at $z \leq 6$.
\citet{huertas-company_galaxy_2024} present $\mathcal{F}_{\mathrm{ET}}$ in $M_{*}$ and $z$ bins.
We fit their results with the analytic expression
\begin{equation}
\label{eq:et_frac}
    \mathcal{F}_{\mathrm{ET}}(M_{*}, z) = \begin{cases}
        \mathcal{F}_{M}(M_{*}) & z < z_{0} \\
        \mathcal{F}_{M}(M_{*}) \left[(1 + z) / (1 + z_{0}) \right]^{-k_{z}} & z \geq z_{0} \, ,
    \end{cases}
\end{equation}
where $\mathcal{F}_{M}(M_{*})$ is a sigmoid of the form
\begin{equation}
    \mathcal{F}_{M}(M_{*}) = \frac{\mathcal{F}_{\mathrm{ET}, 0}}{\mathcal{F}_{\mathrm{ET}, 0} + (1 - \mathcal{F}_{\mathrm{ET}, 0})(M_{*} / 10^{11} \; \mathrm{M}_{\odot})^{- k_{M}}} \, .
\end{equation}
Here, $\mathcal{F}_{\mathrm{ET}, 0}$ is the local galaxy pair fraction at a reference mass of $M_{*} = 10^{11} \; \mathrm{M}_{\odot}$, $z_{0}$ is a characteristic $z$ above which $\mathcal{F}_{\mathrm{ET}}$ decreases as a power-law $\propto z^{k_{z}}$, with power-law slope $k_{z} > 0$, and $k_{M} > 0$ is the sigmoid growth rate.
We find the maximum likelihood values $\mathcal{F}_{\mathrm{ET}, 0} = 0.561$, $z_{0} = 2.428$, $k_{z} = 3.125$, and $k_{M} = 0.492$ provide a good fit to the values of $\mathcal{F}_{\mathrm{ET}}$ presented in \citet{huertas-company_galaxy_2024}.
We also try fitting models with power-law  and sigmoid $z$-dependence to the results of \citet{huertas-company_galaxy_2024}, but find that \autoref{eq:et_frac} provides a better fit than either of those forms.

% We can use the chain rule to express $\phi_{*}$ in terms of the bulge mass as
% \begin{equation}
% \label{eq:bulgemr}
%     \phi_{\mathrm{bulge}} = \frac{d \Phi_{*}}{d \log M_{\mathrm{bulge}}} = \frac{d \Phi_{*}}{d \log M_{*}} \frac{d \log M_{*}}{d \log M_{\mathrm{bulge}}} \, .
% \end{equation}

We next assume a log-linear scaling between $M_{\mathrm{bulge}}$ and SMBH mass, $M_{\mathrm{BH}}$:
\begin{equation}
\label{eq:m-mbulge}
    \log M_{\mathrm{BH}} = \alpha_{*} \log \left(\frac{M_{\mathrm{bulge}}}{10^{11} \; \mathrm{M}_{\odot}}\right) + \beta_{*} \pm \varepsilon_{*} \, ,
\end{equation}
where $\alpha_{*}$ and $\beta_{*}$ are, respectively, the slope and intercept of the $M_{\mathrm{BH}} - M_{\mathrm{bulge}}$ relation, and $\varepsilon_{*}$ is the intrinsic dispersion.
We compute the SMBH mass function, $\phi_{\mathrm{BH}}$, by convolving $\phi_{\mathrm{bulge}}$ with $P(M_{\mathrm{BH}} \vert M_{\mathrm{bulge}})$, which is the probability a galaxy with $M_{\mathrm{bulge}}$ hosts a SMBH with $M_{\mathrm{BH}}$ \citep{marconi_local_2004}:
\begin{equation}
\label{eq:bhmf}
    \phi_{\mathrm{BH}} = \int P(M_{\mathrm{BH}} \vert M_{\mathrm{bulge}}) \phi_{\mathrm{bulge}} d \log M_{\mathrm{bulge}} \, ,
\end{equation}
where\begin{widetext}
\begin{equation}
\label{eq:log-normal}
    P(M_{\mathrm{BH}} \vert M_{\mathrm{bulge}}) = \frac{1}{\sqrt{2 \pi} \varepsilon_{*}} \exp\left\{-\frac{1}{2}\left[\frac{\log M_{\mathrm{BH}} - \beta_{*} - \alpha_{*} \log \left(M_{\mathrm{bulge}} / 10^{11}\; \mathrm{M}_{\odot}\right)}{\varepsilon_{*}}\right]^{2}\right\} \, .
\end{equation}
\end{widetext}

Importantly, our SMBH mass function assumes the same galaxy stellar mass function and mass scalings as our SMBHB mass function, described in the next section.
We therefore take galaxy stellar mass function parameters, $(\phi_{0}, \phi_{1}, M_{0}, \alpha_{0}, \alpha_{1})$, and $M_{\rm BH} - M_{\rm bulge}$ parameters, $(\alpha_{*}, \beta_{*}, \varepsilon_{*})$, from our SMBHB mass function.
This ensures that comparisons between the SMBH and SMBHB mass functions are consistent.

The SMBHB mass function, in turn, is constrained via MCMC using NANOGrav's GWB measurement -- thus our SMBH mass function is also consistent with NANOGrav's GWB measurement.
We assume the galaxy stellar mass functions compiled in \citet{conselice_evolution_2016} and the $M_{\rm BH} - M_{\rm bulge}$ relations compiled in \citet{middleton_no_2018} as astrophysical priors on our MCMC fit.
This spans the range of systematic differences in galaxy stellar mass functions and $M_{\rm BH} - M_{\rm bulge}$ relations.
A description of our SMBHB mass function is provided in the next section, while details of our SMBHB mass function fit and the priors we use are provided in Appendix \ref{sec:mcmc}.

\subsection{Supermassive Black Hole Binary Mass Function}
\label{sec:bhbmf}

    Here we outline how the GWB measurement from \citetalias{agazie_nanograv_2023} provides a constraint on the SMBHB mass function, visualized in the second column at the top of \autoref{fig:duty_cycle_flowchart}, in three steps.
    
    We first model a GWB: we calculate a galaxy pairing rate, then a galaxy merger rate, and finally a SMBHB merger rate. The SMBHB merger rate gives us a GWB amplitude. We then use the measured GWB amplitude from \citetalias{agazie_nanograv_2023} to constrain the SMBHB merger rate via MCMC. Finally, we compute the SMBHB mass function from the constrained SMBHB merger rate and the time to coalescence for SMBHBs in the PTA band.
    
    Specifically, we follow galaxy mergers to SMBHB mergers as in \citetalias{chen_constraining_2019}.
    Starting from $\phi_{*}(M_{*}, z)$ (\autoref{eq:gsmf}) we simultaneously assume a galaxy pair fraction, $\mathcal{F}_{\mathrm{p}}$, and galaxy merger timescale, $\tau_{\mathrm{m}}$, which gives us a galaxy pairing rate, $\dot{\phi}_{*, \mathrm{p}} = d^{3} \Phi_{*,\mathrm{p}} / (d \log M_{*} dq_{*} dt)$.
    This is the differential comoving number density of paired galaxies, $\Phi_{*, \mathrm{p}}$, at pairing redshift $z_{\mathrm{p}}$ per $M_{*}$, galaxy mass ratio, $q_{*}$, and time, $t$.
    We model $\dot{\phi}_{*, \mathrm{p}}$ as
    \begin{equation}
    \label{eq:gmr}
        \dot{\phi}_{*, \mathrm{p}}(M_{*}, z_{\mathrm{p}}, q_{*}) = \phi_{*}(M_{*}, z_{\mathrm{p}}) \dot{\mathcal{F}}_{\mathrm{p}}(M_{*}, z_{\mathrm{p}}, q_{*})\, ,
    \end{equation}
    where $\dot{\mathcal{F}}_{\mathrm{p}} = d^{2} f_{\mathrm{p}} / (dq_{*} dt)$ is the fractional galaxy pairing rate, i.e., the fraction of galaxies that pair per unit time \citep{conselice_early_2006,sesana_systematic_2013,casteels_galaxy_2014,rodriguez-gomez_merger_2015,simon_exploring_2023}.
    We assume that $M_{*}$ is the mass of the more massive galaxy in the merger.
    Further details are given in Appendix \ref{sec:mcmc}.

    We use the galaxy pairing rate to compute the galaxy merger rate, $\dot{\phi}_{*, \mathrm{m}}$, assuming $\dot{\phi}_{*, \mathrm{p}}(M_{*}, z_{\mathrm{p}}, q_{*}) = \dot{\phi}_{*, \mathrm{m}}(M_{*}, z_{\mathrm{m}}, q_{*})$, i.e., the galaxy pairing rate at $z_{\mathrm{p}}$ gives the merger rate at $z_{\mathrm{m}}$, the merger redshift.
    The proper time elapsed between $z_{\mathrm{m}}$ and $z_{\mathrm{p}}$ is given by the galaxy merger timescale, $\tau_{\mathrm{m}}$, which studies have previously constrained using simulations \citep[e.g.][]{kitzbichler_calibration_2008,lotz_effect_2010}.
    The difference between $z_{\mathrm{p}}$ and $z_{\mathrm{m}}$ can thus be found by implicitly solving
    \begin{equation}
    \label{eq:merger_time}
        \int_{z_{\mathrm{m}}}^{z_{\mathrm{p}}} \frac{dt}{dz} dz = \tau_{\mathrm{m}}(M_{*}, z_{\mathrm{p}}, q_{*}) \, ,
    \end{equation}
    where $dt / dz$ is the change in proper time per unit redshift and is given by standard cosmology \citep{hogg_distance_1999}.
    
    To compute the SMBHB merger rate, $\dot{\phi}_{\mathrm{BHB}}$, from the galaxy merger rate, we follow the same steps as in Section \ref{sec:bhmf} to go from a galaxy stellar mass function to a SMBH mass function.
    Specifically, we adopt the mass scalings used to compute $\phi_{\mathrm{BH}}$ from $\phi_{*}$ in Section \ref{sec:bhmf} to compute $\dot{\phi}_{\mathrm{BHB}} = d^{3} \Phi_{\mathrm{BHB}} / (d \log M_{\mathrm{BHB}} \, dq_{\mathrm{BHB}} \, dt)$ from $\dot{\phi}_{*, \mathrm{m}}$.
    As in \citetalias{chen_constraining_2019} and many other works \citep[e.g.,][]{sesana_systematic_2013,sesana_selection_2016,chen_probing_2017}, the SMBH merger timescale is equal to the galaxy merger timescale.
    To test this assumption we also consider a delay equal to the dynamical friction timescale \citep[e.g.,][]{binney_galactic_2008,mingarelli_local_2017}.
    We find this has negligible impact on our results and that the SMBHB merger rate is not particularly sensitive to the time delay between galaxy pairing and SMBHB merger.

    We next place constraints on $\dot{\phi}_{\mathrm{BHB}}$ using the recent GWB characteristic strain measurement from \citetalias{agazie_nanograv_2023}.
    The characteristic strain can be computed from $\dot{\phi}_{\mathrm{BHB}}$ as \citep{phinney_practical_2001,sesana_stochastic_2008}
    \begin{align}
    \label{eq:gwb}
    \begin{split}
         h_{c}^{2}(f_{\mathrm{GW}}) &= \frac{4}{3 \pi}\frac{1}{f_{\mathrm{GW}}^{4 / 3}} \iiint \dot{\phi}_{\rm BHB} \frac{dt}{dz} \frac{M_{\mathrm{BHB}}^{5 / 3}}{(1 + z_{\mathrm{m}})^{1 /3}} \\ & \qquad \times \frac{q_{\mathrm{BHB}}}{(1 + q_{\mathrm{BHB}})^{2}} d\log M_{\rm BHB} dz_{\mathrm{m}} dq_{\mathrm{BHB}} \, .
    \end{split}
    \end{align}
    Measurements of the characteristic strain are generally quoted at $A_{\mathrm{GWB}} = h_{c}(f_{\mathrm{GW}} = 1 \; \mathrm{yr}^{-1})$, and we know from \citetalias{agazie_nanograv_2023} that $A_{\mathrm{GWB}} = \left(2.4^{+0.7}_{-0.6}\right) \times 10^{-15}$.
    We use this result to constrain $\dot{\phi}_{\mathrm{BHB}}$ with MCMC via \autoref{eq:gwb}, with a reduced $\chi^{2}$ of $0.997$.
    Details of our fit and the resulting posterior distribution are given in Appendix \ref{sec:mcmc}.
    
    Finally, we multiply $\dot{\phi}_{\mathrm{BHB}}$ by the time to coalescence for a PTA band SMBHB to compute the SMBHB mass function, $\phi_{\mathrm{BHB}}$.
    Specifically, this is the mass function of SMBHBs emitting in the observer frame PTA band.
    For circular SMBHBs, the time to coalescence is
    \begin{equation}
        T_{c} = \frac{5}{256} M_{\rm BHB}^{-5 / 3} \frac{(1 + q_{\rm BHB})^{2}}{q_{\rm BHB}} \left[\pi f_{\mathrm{GW}} (1 + z) \right]^{-8 / 3} \, ,
    \end{equation}
    where $f_{\mathrm{GW}} = 10^{-9} \; \mathrm{Hz}$ is the lowest frequency in the PTA band.

\subsection{Quasar Mass Function}
\label{sec:qmf}

    Next we calculate a SMBHB occupation fraction for quasars, \autoref{fig:duty_cycle_flowchart}.
    For this we need a quasar mass function, which we calculate here, and a binary quasar mass function, Section \ref{sec:bqmf}.
    
    To calculate the fraction of quasars hosting SMBHBs we first need to model the quasar population.
    We assume the ``convolutional" quasar mass function for type-1 AGN from \citet[S20 hereafter]{shen_bolometric_2020}.
    We start from a bolometric quasar luminosity function (QLF) based on compiled observations in the IR, B, UV, soft, and hard X-ray bands.
    We then calculate the quasar mass function by convolving the QLF with an empirical Eddington ratio distribution function (ERDF) based on X-ray selected AGN at $z \sim 1.4$ (\citealp{nobuta_black_2012}, \citetalias{shen_bolometric_2020}).
    \citetalias{shen_bolometric_2020} reports that this choice of ERDF \citep{nobuta_black_2012} reproduces the observed mass distribution of SMBHs in the local universe \citep{marconi_local_2004,shankar_selfconsistent_2009,vika_millennium_2009}, as well as the observed mass distribution of type-1 AGN (\citealp{kelly_demographics_2013}, \citetalias{shen_bolometric_2020}).
    
\subsection{Binary Quasar Mass Function}
\label{sec:bqmf}

The final ingredient for calculating a SMBHB occupation fraction for quasars is a binary quasar mass function.
To compute this we first construct the binary QLF using the CRTS sample.
We then convolve the binary QLF with the ERDF to compute a binary quasar mass function.
The details are as follows.

The binary QLF, $\phi_{\mathrm{BQ}}(L, z) = d \Phi_{\mathrm{BQ}} / d \log L$, is the differential number density of binary quasars per $\log L$, where $L$ is bolometric luminosity.
Similar to the galaxy stellar mass function in \autoref{eq:gsmf}, we model the binary QLF as a $z$ evolving Schechter function,
\begin{align}
\begin{split}
    \label{eq:bqlf}
   \phi_{\mathrm{BQ}}(L, z) &= \Phi^{\mathrm{BQ}}_{*}(z)\left(\frac{L}{L_{*}(z)}\right)^{1 + \alpha^{\mathrm{BQ}}_{*}(z)}  \\
   & \qquad \times \exp\left(-\frac{L}{L_{*}(z)}\right) ,
\end{split}
\end{align}
which we find reproduces the expected rarity of high mass -- and therefore high luminosity -- binary quasars.
We take the $z$ evolving parameters as
\begin{align}
    \log \Phi^{\mathrm{BQ}}_{*}(z) &= \phi^{\mathrm{BQ}}_{0} + \phi^{\mathrm{BQ}}_{1}(z) \label{eq:bq_norm}\\
    L_{*}(z) &= \frac{2 L_{0}}{\left(\frac{1 + z}{1 + z_{\mathrm{ref}}}\right)^{\gamma_{1}} + \left(\frac{1 + z}{1 + z_{\mathrm{ref}}}\right)^{\gamma_{2}}} \\
    \alpha^{\mathrm{BQ}}_{*}(z) &= \alpha^{\mathrm{BQ}}_{0} + \alpha^{\mathrm{BQ}}_{1}(z)
\end{align}
where the linear evolution of $\Phi^{\mathrm{BQ}}_{*}(z)$ and $\alpha^{\mathrm{BQ}}_{*}(z)$ are analogous to their corresponding parameters in \autoref{eq:gsmf}.
$L_{*}(z)$ takes the form of a double power law, mimicking the form of the break luminosity in our assumed QLF \citepalias{shen_bolometric_2020}.
We find this parameterization of $L_{*}(z)$, with $z_{\mathrm{ref}} = 0.75$, is necessary to ensure the binary quasar mass function is less than both the quasar mass function and the SMBHB mass function at all $M_{\mathrm{BHB}}$ and $z$.
Finally, we convolve the constrained binary QLF with the ERDF to compute the binary quasar mass function, similar to our computation of the quasar mass function.

\subsection{Occupation Fractions}
\label{sec:fractions}

The binary occupation fraction, $\mathcal{F}_{\mathrm{BHB}} = \phi_{\mathrm{BHB}} / \phi_{\mathrm{BH}}$, characterizes the fraction of massive galaxies hosting an SMBHB.
We can similarly calculate the fraction of quasars that we expect to host SMBHBs, $\mathcal{F}_{\mathrm{BHB \vert QSO}} = \phi_{\mathrm{BQ}} / \phi_{\mathrm{QSO}}$.
By comparing $\mathcal{F}_{\mathrm{BHB \vert QSO}}$ to $\mathcal{F}_{\mathrm{BHB}}$ we can quantify whether or not quasars signpost SMBHBs --- the final step in \autoref{fig:duty_cycle_flowchart}.

\section{Results}
\label{sec:results}

Here we present constraints on the binary quasar population.
We start by using the SMBHB mass function and the quasar mass function to constrain the number of genuine SMBHBs in the CRTS sample.
Following this we constrain the binary QLF, which we use in turn to compute the binary quasar mass function.
Finally, we compute and compare the binary occupation fraction for galaxies and for quasars.

    \subsection{Upper Limits on Genuine Binaries}
    \label{sec:upper}
    
    Several studies claim that the CRTS binary candidates likely include false positives \citep{vaughan_false_2016,sesana_testing_2018,kelley_massive_2019,witt_quasars_2022,davis_reliable_2024} -- we further investigate this claim here.
    We use the fact that, by definition, there cannot possibly be more binary quasars than either SMBHBs or quasars.
    Consequently, constraints on SMBHBs and quasars also imply constraints on binary quasars.
    
    We start our investigation by estimating the bolometric luminosities of the CRTS candidates for a consistent comparison with the bolometric QLF.
    We first use the $V$-band magnitude of each candidate to approximate optical luminosities \citep{tachibana_deep_2020}.
    We include empirical dust corrections \citep{odonnell_vdependent_1994} and $K$ corrections \citep{hogg_correction_2002} calculated consistently with \citetalias{shen_bolometric_2020}.
    We then calculate the bolometric luminosity of each candidate using optical-bolometric corrections from \citetalias{shen_bolometric_2020}.
    We assume that dispersion in the bolometric corrections are small in the luminosity range we consider ($\gtrsim 10^{45} \; \mathrm{erg}\;\mathrm{s}^{-1}$).
    
    We next bin the CRTS candidates, $i = 1, ..., 87$, over bolometric luminosity and $z$.
    We place an upper limit on the expected number of genuine binary quasars by assuming that \textit{all} quasars are binaries.
    In each bin $j$ we integrate the QLF over luminosity and $z$ to calculate the maximum expected number of binary quasars, $N^{\mathrm{QSO}}_{\mathrm{BQ}, \max, j}$, accounting for CRTS selection effects, including the limiting magnitude and effective completeness of CRTS (Section~\ref{sec:complete}).
    We include both Poisson and Monte Carlo model uncertainties.
    
    We similarly compute upper limits on binary quasars using the SMBHB mass function by assuming that all binaries have associated quasar activity (i.e., there cannot possibly be more binary quasars than SMBHBs).
    We infer the maximum binary QLF implied by the GWB by deconvolving the SMBHB mass function with the ERDF.
    As with the empirical QLF used above, we then integrate over luminosity and $z$ to calculate the maximum expected number of binary quasars, $N^{\mathrm{GWB}}_{\mathrm{BQ}, \max}$ in each bin.
    For each bin we then conservatively take the maximum number of binary quasars to be the smaller value predicted by either quasars or the GWB, i.e., $N_{\mathrm{BQ}, \max, j} = \min\left\{N^{\mathrm{QSO}}_{\mathrm{BQ}, \max, j}, N^{\mathrm{GWB}}_{\mathrm{BQ}, \max, j}\right\}$.

\begin{figure}
    \centering
    \includegraphics[width=\columnwidth]{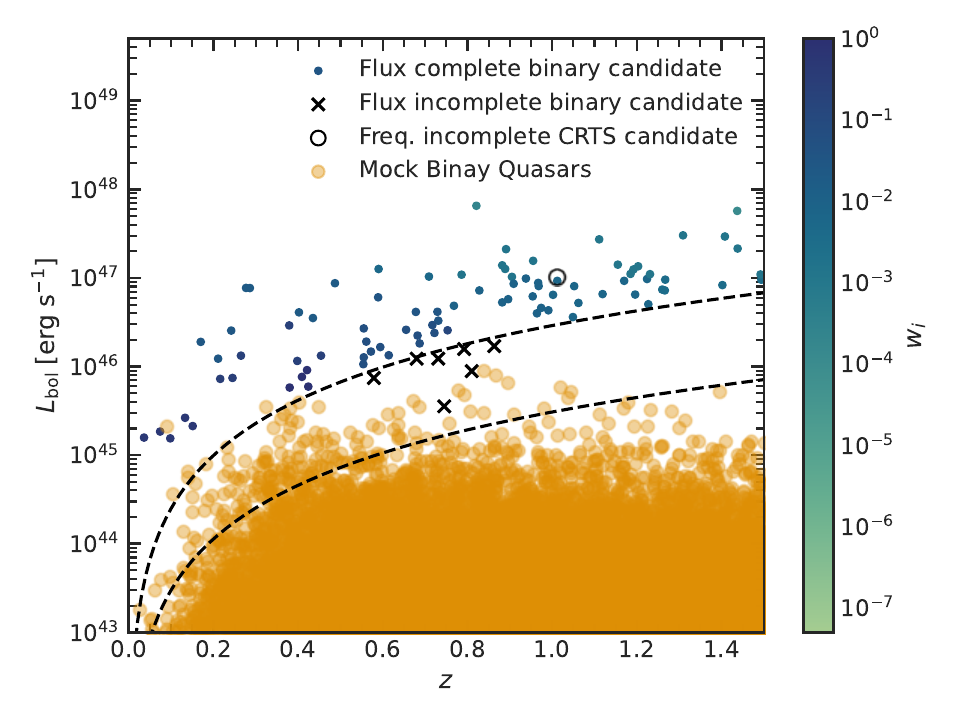}
    \caption{
    Binary quasar candidates from CRTS (variable color dots, with color bar denoting the expected likelihood that each candidate is genuine), and a parent sample of mock binary quasars (orange dots). We also plot the flux completeness limit (dashed line) and maximum depth (dash-dotted line) of CRTS. Our analysis does not consider flux incomplete CRTS candidates (black ``x"s) or SDSS J113916.47+254412.6 (black ``o"), whose periodicity is below the formal cutoff from \citetalias{graham_systematic_2015}. Vera C. Rubin LSST will observe quasars more than an order of magnitude fainter than the faintest quasars seen by CRTS (dotted line).}
    \label{fig:mock_quasar_catalog}
\end{figure}

    Finally, we combine these upper limits to estimate the maximum number of genuine binaries in CRTS.
    In each bin we compare the number of binary quasar candidates observed by CRTS, $N_{\mathrm{CRTS}, j}$ to $N_{\mathrm{BQ}, \max, j}$.
    We then weight each candidate according to $w_{i} = \min\left\{1, N_{\mathrm{BQ}, \max, i \in j} / N_{\mathrm{CRTS}, i \in j}\right\}$ so that the weighted sum of candidates in each bin is at most $N_{\mathrm{BQ}, \max, j}$.
    For example, if we find $4$ CRTS candidates in a bin where we expect there to be at most only $2$, each CRTS candidate is assigned a weight of $0.5$.
    This reflects our expectation that $\leq 50 \%$ of those candidates could be genuine binaries.
    The maximum expected number of SMBHBs in CRTS is then $N_{\mathrm{BQ}, \max} = \sum_{i} w_{i}$.
    
    We find $\lesssim 8$ CRTS candidates are likely to be genuine, at $95 \%$ confidence.
    For comparison, \citet{kelley_massive_2019} found $\lesssim 5$ CRTS candidates may be genuine using hydrodynamic simulations.
    Interestingly, the number of binaries in a given frequency interval scales with observation time as $T_{\rm obs}^{8 / 3}$.
    This implies that continued time domain quasar observations will reveal an increasing number of SMBHBs at lower $f_{\mathrm{GW}}$ than is accessible to CRTS.
    For example, if all CRTS quasar light curves are extended by a  decade -- such as by Vera C. Rubin LSST \citep{ivezic_lsst_2019} -- we would expect up to $8 \times (19 \; \mathrm{yr})^{8 / 3} / (9 \; \mathrm{yr})^{8 / 3} \approx 59$ genuine binaries to manifest as periodic light curves.
    
    We also find that GWB constraints on the binary quasar population predict at most $\mathcal{O}(10)$ PTA-band SMBHBs above $10^{47} \; \mathrm{erg} \; \mathrm{s}^{-1}$.
    This suggests that the brightest binary quasar candidates are the least likely to host a genuine binary, as these candidates would be associated with the most massive -- and therefore rarest -- SMBHBs.
    Vera C. Rubin LSST will observe much fainter quasars than CRTS (\autoref{fig:mock_quasar_catalog}) -- it will therefore be crucial for constraining the faint binary quasar population.

\subsection{Binary Quasar Luminosity Function}
\label{sec:bqlf}

Here we constrain the binary QLF needed to compute the binary quasar mass function.
We find that modelling the knee and faint-end slope of the binary QLF is important for its correct reconstruction.
However most of the CRTS sample is bright, with $L \geq 10^{46} \; \mathrm{erg} \; \mathrm{s}^{-1}$, \autoref{fig:mock_quasar_catalog}.
We thus estimate the distribution of low luminosity binary quasars using limits inferred from quasar observations and the GWB.
We then generate a parent sample of binary quasars from this distribution that is matched to $N_{\mathrm{BQ}, \max}$, and which is complete to $z = 1.5$ above $10^{43} \; \mathrm{erg} \; \mathrm{s}^{-1}$.
We also correct for the effective completeness (Section~\ref{sec:complete}) and $f_{\mathrm{GW}}$ coverage (Section~\ref{sec:pta}) of CRTS so that the parent sample represents a full-sky population of binary quasars with $1 \; \mathrm{nHz} \leq f_{\mathrm{GW}} \leq 100 \; \mathrm{nHz}$.
We finally use this parent sample to construct the binary QLF upper limit.

We first extrapolate the low luminosity behavior of the binary QLF from the maximum number of binary quasars inferred by quasars and the GWB.
We compute the combined maximum binary QLF as $\phi_{\mathrm{BQ}, \max}(L, z) = \min\left\{\phi_{\mathrm{QSO}}(L, z), \phi_{\mathrm{BHB}}(L, z)\right\}$, where $\phi_{\mathrm{QSO}}(L, z)$ is the empirical QLF \citepalias{shen_bolometric_2020}, and $\phi_{\mathrm{BHB}}(L, z)$ is the maximum binary QLF inferred from SMBHBs, described in Section \ref{sec:upper}.

We next generate a parent sample of $3.7 \times 10^{4}$ binary quasars that is complete to $z = 1.5$ above $L = 10^{43} \; \mathrm{erg} \; \mathrm{s}^{-1}$, \autoref{fig:mock_quasar_catalog}.
We draw samples from $\phi_{\mathrm{BQ}, \max}(L, z)$, statistically matching the number of samples drawn to $N_{\mathrm{BQ}, \max}$ in the flux complete volume of CRTS.

\begin{figure}
    \centering
    \includegraphics[width=\columnwidth]{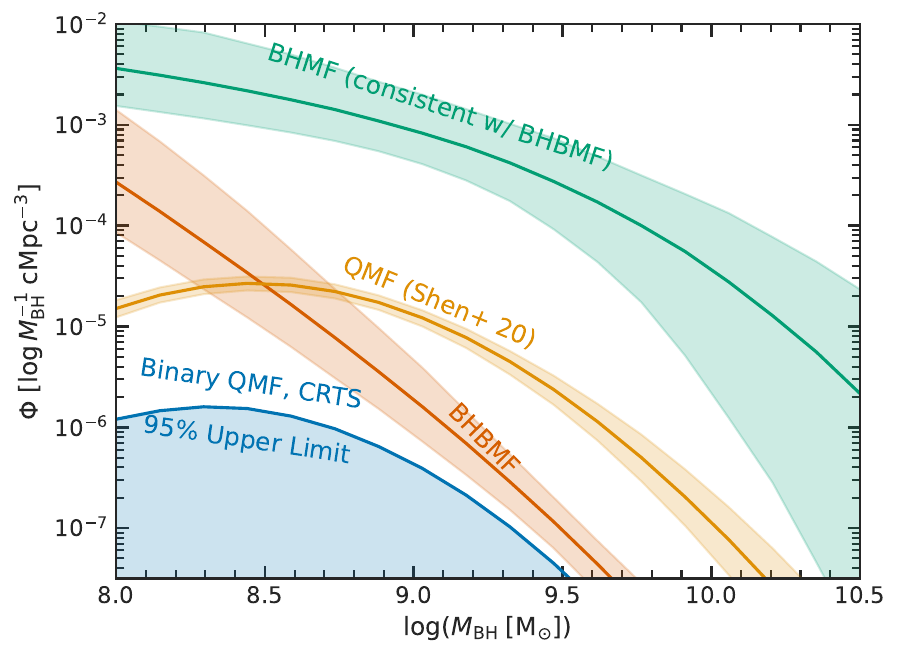}
    \caption{Mass functions from Section \ref{sec:mfs}. Here we show the SMBH mass functions we consider: the $z$-integrated SMBH mass function (green, Section \ref{sec:bhmf}), SMBHB mass function (red, Section \ref{sec:bhbmf}), quasar mass function (orange, Section \ref{sec:qmf}), and binary quasar mass function 95 \% upper limit (blue, Section \ref{sec:bqmf}).
    For mass functions other than the binary quasar upper limit the shaded regions denote $95 \%$ confidence intervals.
    The low-mass turnover seen in both the quasar and binary quasar mass functions is a result of considering only $L \gtrsim 10^{45} \; \mathrm{erg} \; \mathrm{s}^{-1}$.
    The broad uncertainties in the SMBH and SMBHB mass functions are primarily due to uncertainties in the $M_{\mathrm{BH}} - M_{\mathrm{bulge}}$ relation.
    }
    \label{fig:mass_functions}
\end{figure}

We then fit $\phi_{\mathrm{BQ}}(L, z)$ using the STY method proposed by \citet{sandage_velocity_1979}. The STY method is a parametric maximum likelihood technique for fitting a luminosity function to a population of astrophysical objects.
The differential number of binary quasars per $\log L$ and $z$ is $\mathcal{N}_{\mathrm{BQ}}(L, z) = \phi_{\mathrm{BQ}}(L, z) dV / dz$, where $dV / dz$ is the differential change in comoving volume per unit $z$.
It thus follows that the probability, $p_{i}$, of observing object $i$ in a sample of binary quasars is \citep{sandage_velocity_1979,weigel_stellar_2016}
\begin{equation}
\label{eq:indiv_prob}
    p_{i}(L_{\mathrm{bol}, i}, z_{i}) = \frac{\mathcal{N}_{\rm BQ}(L_{\mathrm{bol}, i}, z_{i})}{N_{\mathrm{BQ}}} \, ,
\end{equation}
where $N_{\mathrm{BQ}} = \int_{z_{\min}}^{z_{\max}} \int_{L_{\min}}^{L_{\max}} \mathcal{N}_{\rm BQ}(L, z) d \log L \; dz$ is the total number of $10^{45} \leq L \leq 10^{48} \; \mathrm{erg} \; \mathrm{s}^{-1}$ binary quasars in $0 \leq z \leq 1.5$, matching the CRTS sample.
We use the MCMC sampler \texttt{pymc} \citep{wiecki_pymcdevs_2023} to maximize the log likelihood, $\ln \mathcal{L} = \sum_{i} \ln p_{i}$, of observing all objects in the sample.
Importantly, \autoref{eq:indiv_prob} cannot directly constrain $\phi_{0}^{\mathrm{BQ}}$, the normalization of $\phi_{\mathrm{BQ}}$ (Equations~\ref{eq:bqlf} and \ref{eq:bq_norm}).
We instead constrain $\phi_{0}^{\mathrm{BQ}}$ by computing \begin{equation}
10^{\phi_{0}^{\mathrm{BQ}}} = \frac{N_{\mathrm{BQ}}}{\Omega_{\mathrm{CRTS}} S_{\mathrm{GW}} \int_{z_{\min}}^{z_{\max}} \int_{L_{\min}}^{L_{\max}} \mathcal{N}'_{\rm BQ}(L, z) d \log L \; dz} \, ,
\end{equation} where $\Omega_{\mathrm{CRTS}} = 0.33$ is the effective sky coverage of CRTS (Section~\ref{sec:complete}), $S_{\mathrm{GW}} = 0.002$ is the PTA band coverage of CRTS (Section~\ref{sec:pta}), and where $\mathcal{N}'_{\rm BQ} = \mathcal{N}_{\rm BQ} / 10^{\phi_{0}^{\mathrm{BQ}}}$ \citep[cf.][]{weigel_stellar_2016}.
Finally, we compute the binary quasar mass function, \autoref{fig:mass_functions}.

We verify that our results are consistent with observations in two ways.
Firstly, we compute the GWB amplitude using only binary quasars, and find this to be at most $6.1 \times 10^{-16}$, almost four times less than the GWB measurement reported in~\citetalias{agazie_nanograv_2023}.
We next integrate our binary quasar mass function in the NANOGrav 15 year CW volume -- that is the $f_{\mathrm{GW}}$-dependent volume accessible to NANOGrav's 15 year CW search~\citep{agazie_nanograv_2023e}.
We find $\ll 1$ quasar-based SMBHB system is expected to be detectable, consistent with the current CW non-detections.

\subsection{Binary Fractions of Quasars and Galaxies}

\begin{figure}
    \centering
    \includegraphics[width=\columnwidth]{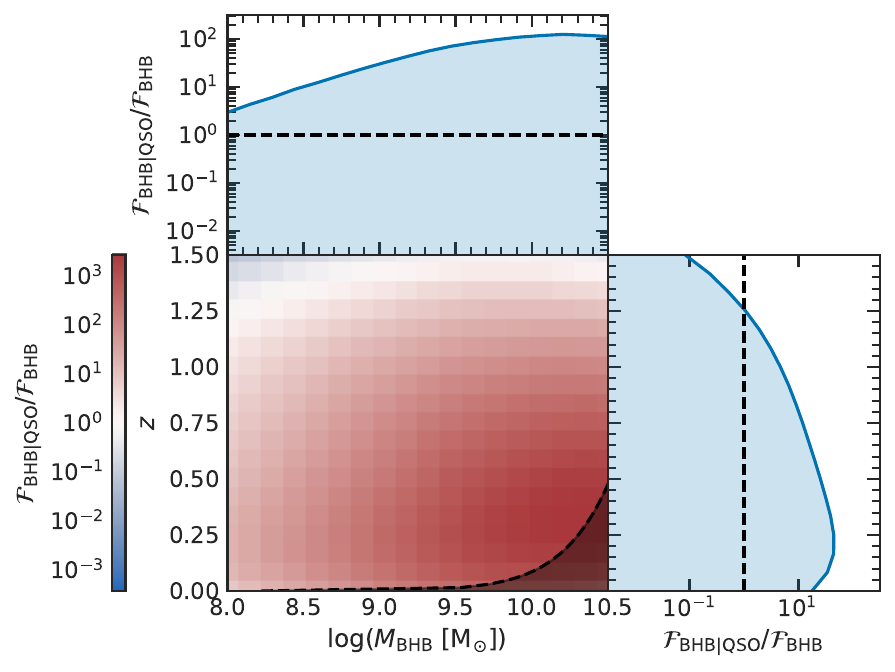}
    \caption{SMBHB occupation fractions of quasars vs. massive galaxies, shown centrally as a function of $M_{\rm BHB}$ and $z$ simultaneously, and in the panels as a function of each individually.
    Shaded regions in the side panels show the $95 \%$ upper limit, while the dashed lines show the case where $\mathcal{F}_{\mathrm{BHB \vert QSO}} = \mathcal{F}_{\mathrm{BHB}}$.
    The fraction of quasars hosting binaries is larger than the fraction of galaxies hosting binaries over all masses, indicating that quasars may signpost binaries at any mass.
    Likewise, quasars may signpost binaries at any $z$ except for in a local volume.
    We also show the region of $M_{\rm BH} - z$ space that is detectable by the NANOGrav 15 year dataset at its most sensitive frequency \citep[black dashed line,][]{agazie_nanograv_2023e}, and is therefore ruled out.
    }
    \label{fig:occ_frac_comparison}
\end{figure}

We calculate the average binary occupation fraction by integrating each mass function over $10^8 \leq M_{\rm BH} \leq 10^{10.5} \; M_\odot$ and $0 \leq z \leq 1.5$, giving us the comoving number densities of SMBHs, SMBHBs, quasars, and binary quasars, \autoref{fig:duty_cycle_flowchart}.
We find the average binary occupation fraction for massive galaxies is $\bar{\mathcal{F}}_{\rm BHB} \approx 2.6^{+4.8}_{-1.8} \%$, where the error bars denote the $95 \%$ confidence interval.
Repeating this calculation with the quasar mass function yields the $95 \%$ upper limit $\bar{\mathcal{F}}_{\rm BHB \vert QSO} \lesssim 5 \%$.
Considering the range of uncertainties in $\bar{\mathcal{F}}_{\rm BHB}$ and $\bar{\mathcal{F}}_{\rm BHB \vert QSO}$, this indicates that quasars are up to five times more likely to host a SMBHB than a random galaxy is.
This upper limit corresponds to a universe with $\mathcal{F}_{\mathrm{BHB}}$ near its lower limit of $0.4 \%$, such that the fraction of galaxies hosting a SMBHB is small compared to the fraction of quasars hosting a SMBHB.
Interestingly, such a universe would necessarily have binaries which are more massive than those in a universe with a larger $\bar{\mathcal{F}}_{\rm BHB}$ and the same $A_{\mathrm{GWB}}$ \citep{casey-clyde_quasarbased_2022}.
We show $\mathcal{F}_{\rm BHB \vert QSO} / \mathcal{F}_{\rm BHB}$ as a function of both mass and $z$ in \autoref{fig:occ_frac_comparison}.

\vspace{1cm}
\section{Summary and Conclusion} \label{sec:conclusion}

We have investigated whether or not quasars can preferentially host SMBHBs compared to random galaxies.
Our investigation combines information from quasar observations \citepalias{shen_bolometric_2020} and the GWB \citepalias{agazie_nanograv_2023} to constrain the binary quasar population.
This enables us to compute upper limits on the binary occupation fraction of quasars, which we then compare to the binary occupation fraction of galaxies.

We use the CRTS catalog \citepalias{graham_systematic_2015} to derive an upper limit on binary quasars.
We constrain the number of genuine CRTS binaries using quasar observations \citepalias{shen_bolometric_2020} and the GWB measurement from \citetalias{agazie_nanograv_2023}.
We find that $\lesssim 8$ of the CRTS candidates may be SMBHBs, in agreement with \citet{kelley_massive_2019}.
The future detection of a binary in a quasar will enable direct constraints on the binary quasar population, rather than an upper limit.

Interestingly, we find the most luminous AGN in CRTS are the least likely to host a genuine binary.
This is a consequence of the expected rarity of high mass SMBHBs \citep{casey-clyde_quasarbased_2022}, which implies high luminosity binary quasars should also be rare.
In the next few years we expect CW observations will start definitively confirming/rejecting nearby SMBHB candidates \citep{mingarelli_local_2017,xin_multimessenger_2021}.
If any binary quasar candidates can be confirmed, the combined multi-messenger information may yield significant insights to how the binary's orbit imprints on the observed quasar light-curve \citep[e.g.][]{dorazio_accretion_2013,farris_binary_2014,miranda_viscous_2017,munoz_circumbinary_2020}, in addition to the GW signal.

CRTS is too shallow to effectively probe quasars with $L \lesssim 10^{45} \; \mathrm{erg} \; \mathrm{s}^{-1}$ in the volume $z \leq 1.5$.
We expect Vera Rubin LSST to be complete to $L \gtrsim 10^{44} \; \mathrm{erg} \; \mathrm{s}^{-1}$ in this volume, \autoref{fig:mock_quasar_catalog}, placing stricter constraints on the SMBHB population.
Since the number of binaries at a given frequency scales with observation time as $T_{\rm obs}^{8 / 3}$, continued time-domain observations of quasars will reveal an increasing number of binary quasars with longer orbital periods.
We therefore recommend time-domain quasar monitoring to continue as long as possible.
Upcoming time-domain surveys such as Vera C. Rubin LSST will be crucial for improving constraints on the binary quasar population.

Using our new multimessenger technique we found that, at 95\% confidence, $2.6^{+4.8}_{-1.8} \%$ of massive galaxies, and $\lesssim 5 \%$ of quasars, host a SMBHB.
Our SMBH mass function is fully consistent with our SMBHB mass function by construction, as we build our SMBH mass function using only models and assumptions present in our SMBHB mass function.
Our predicted SMBHB occupation fraction is thus self-consistent.
This self-consistency is a feature of any SMBHB model derived from galaxy major-mergers (\citealp{sesana_stochastic_2008,sesana_systematic_2013,chen_probing_2017}, \citetalias{chen_constraining_2019}), though we are the first to use and highlight this fact.

Finally, comparing the binary occupation fraction of galaxies to the occupation fraction of quasars, we thus find that quasars are up to five times more likely to host a binary than a random galaxy.
Thus, despite the fact that only $8$ of the $111$ binary quasar candidates are likely to be genuine, quasars are not ruled out as preferential SMBHB hosts.
Thus, targeted CW searches of quasars may be more likely to find genuine SMBHBs than searches on a random selection of galaxies.
Moreover, confirming/rejecting binary quasar candidates (e.g., the CRTS candidate catalog) as genuine will be crucial for improving constraints on $\mathcal{F}_{\mathrm{BHB | QSO}}$.
Indeed, the possibility of no genuine SMBHBs among the CRTS candidates can only be ruled out by a CW detection.

Interestingly, a more massive SMBHB population could imply a higher upper limit on $\mathcal{F}_{\mathrm{BHB | QSO}} / \mathcal{F}_{\mathrm{BHB}}$.
This is due to the fact that a more massive SMBHB population requires fewer binaries (i.e., a smaller $\mathcal{F}_{\mathrm{BHB}}$) to produce the same GWB (\autoref{eq:gwb}, \citealt{casey-clyde_quasarbased_2022}).
Recent observations from the James Webb Space Telescope suggest a population of overmassive SMBHs relative to their host galaxies at high redshift \citep[e.g.,][]{li_tip_2024}.

To summarize, using current observations of quasars and the GWB we constrain the number of genuine binary quasars in the CRTS binary candidate catalog.
We find the majority of the catalog are likely to be false positives.
We then determine if an excess of SMBHBs among the quasar population can be ruled out, given that only a few CRTS candidates may be genuine.
Nonetheless, we find that SMBHBs are up to five times more likely to be found in quasars than in random massive galaxies.
We thus conclude that if even just a few of the CRTS candidates are genuine, quasars may signpost SMBHBs emitting nHz GWs. Quasars should therefore be prioritized as targets for CW searches with PTAs, which will be crucial for constraining the binary quasar population.

\vspace{\baselineskip}
%\begin{minipage}{\columnwidth}
\noindent The authors thank the referee for their keen interest in this work and insightful comments, which helped make this work more robust. The authors also thank Matthew Graham, Meg Davis, Deborah Good, Bjorn Larsen, London Wilson, Jessie Runnoe, Caitlin Witt, Luke Kelley, Joe Lazio, Joe Simon, Jennifer Wallace, Skylar Wright, Nikko Cleri, and Daniel Sniffin. This research was supported in part by the National Science Foundation under Grants NSF PHY-2020265, and AST-2106552. The Flatiron Institute is supported by the Simons Foundation. JRT acknowledges support from NSF grants CAREER-1945546, AST-2009539, and AST-2108668.
%\end{minipage}

\software{
    astroquery \citep{ginsburg_astroquery_2019},
    astropy \citep{astropycollaboration_astropy_2013,astropycollaboration_astropy_2018},
    corner \citep{foreman-mackey_cornerpy_2016},
    emcee \citep{foreman-mackey_emcee_2013},
    jupyter \citep{kluyver_jupyter_2016},
    matplotlib \citep{hunter_matplotlib_2007},
    nHzGWs \citep{mingarelli_chiaramingarelli_2017},
    numpy \citep{harris_array_2020},
    pandas \citep{mckinney_data_2010,thepandasdevelopmentteam_pandasdev_2022},
    pymc \citep{wiecki_pymcdevs_2023},
    scipy \citep{virtanen_scipy_2020},
    seaborn \citep{waskom_seaborn_2021}
}

\appendix
\section{Gravitational Wave Background Constraints}
\label{sec:mcmc}

\begin{figure*}[tbh]
    \centering
     \gridline{\fig{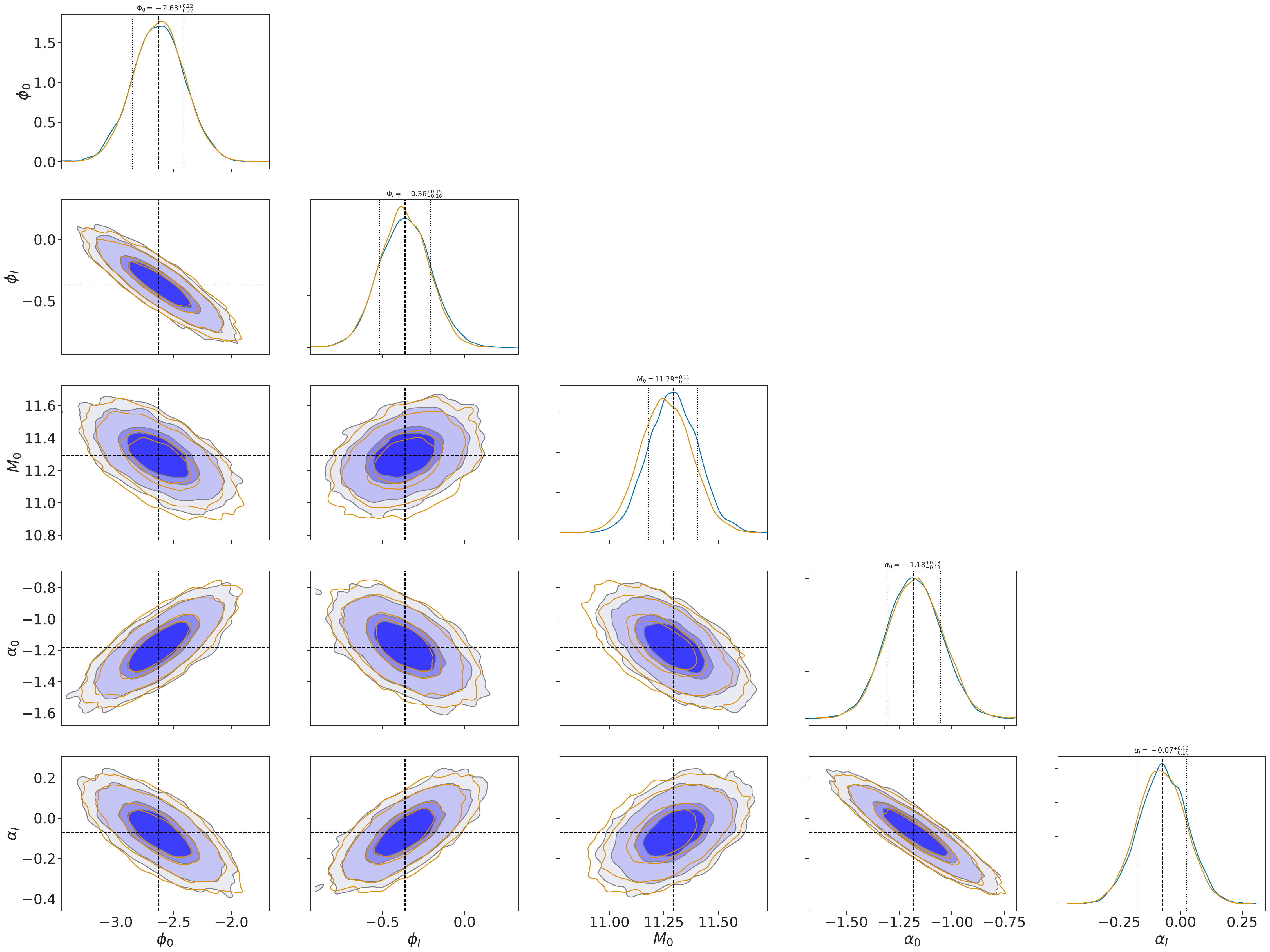}{0.45\textwidth}{(a) GSMF parameters.}
              \fig{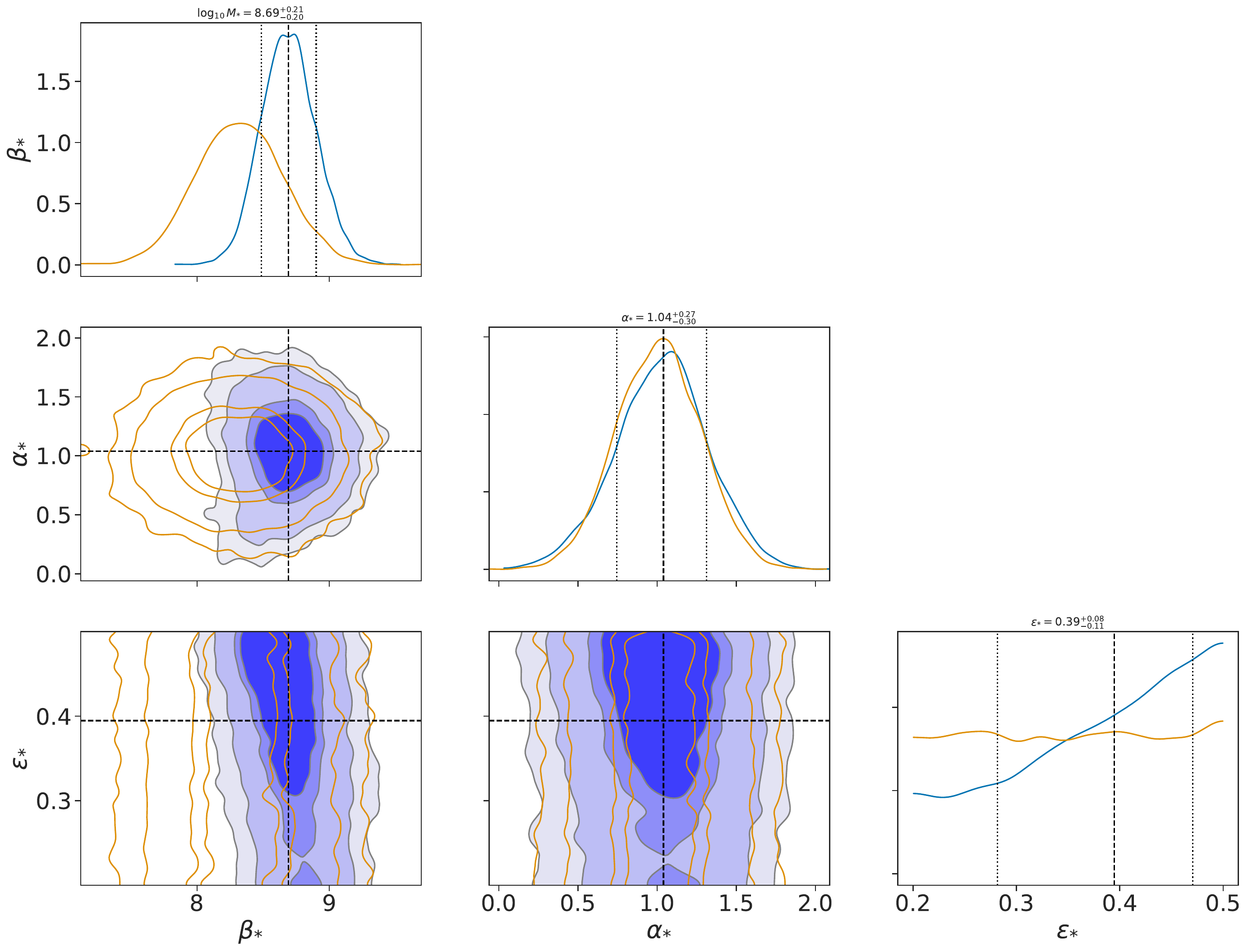}{0.45\textwidth}{(b) $M_{\mathrm{BH}} - M_{\mathrm{bulge}}$ parameters.}}
    \gridline{\fig{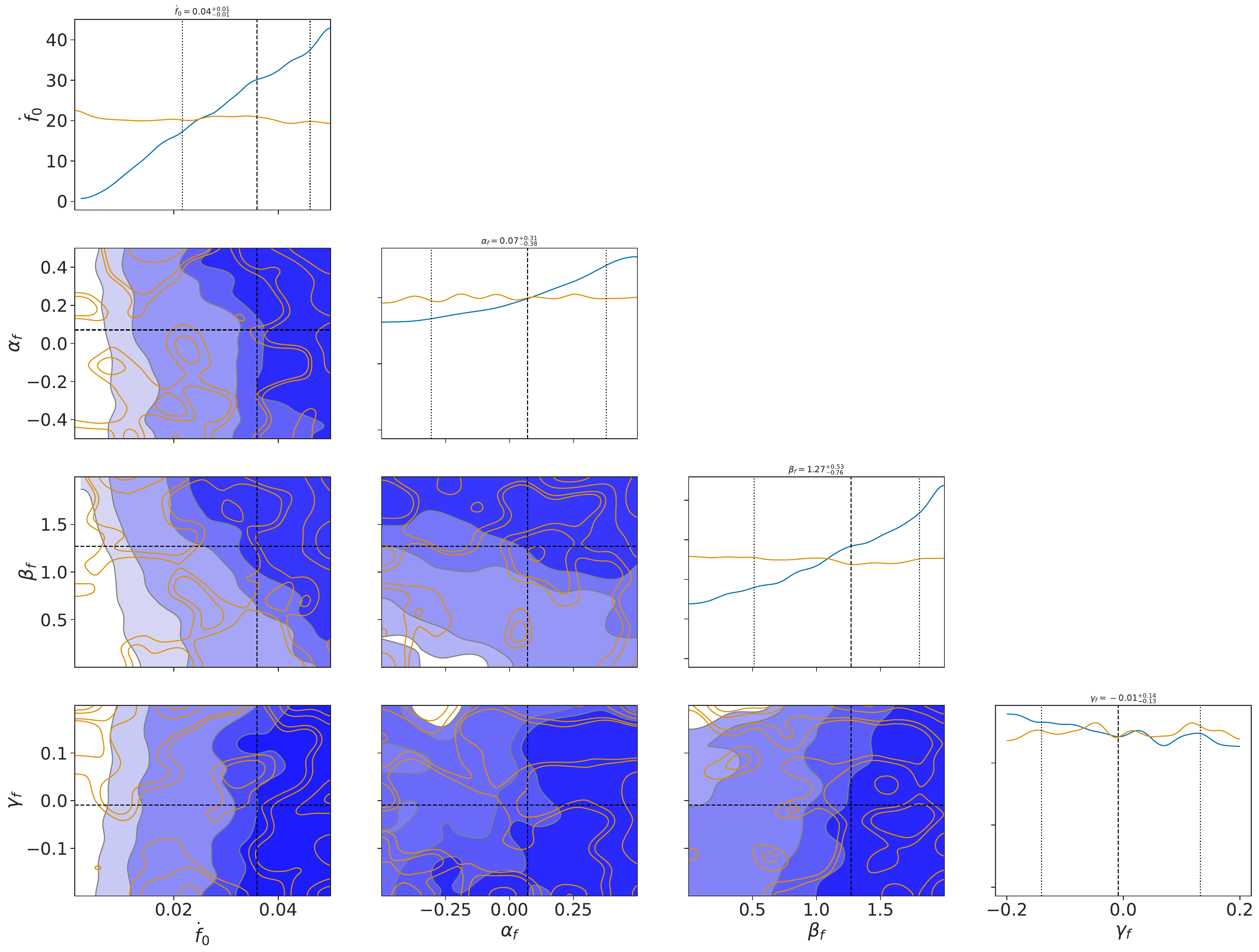}{0.45\textwidth}{(c) Fractional pairing rate parameters.}
              \fig{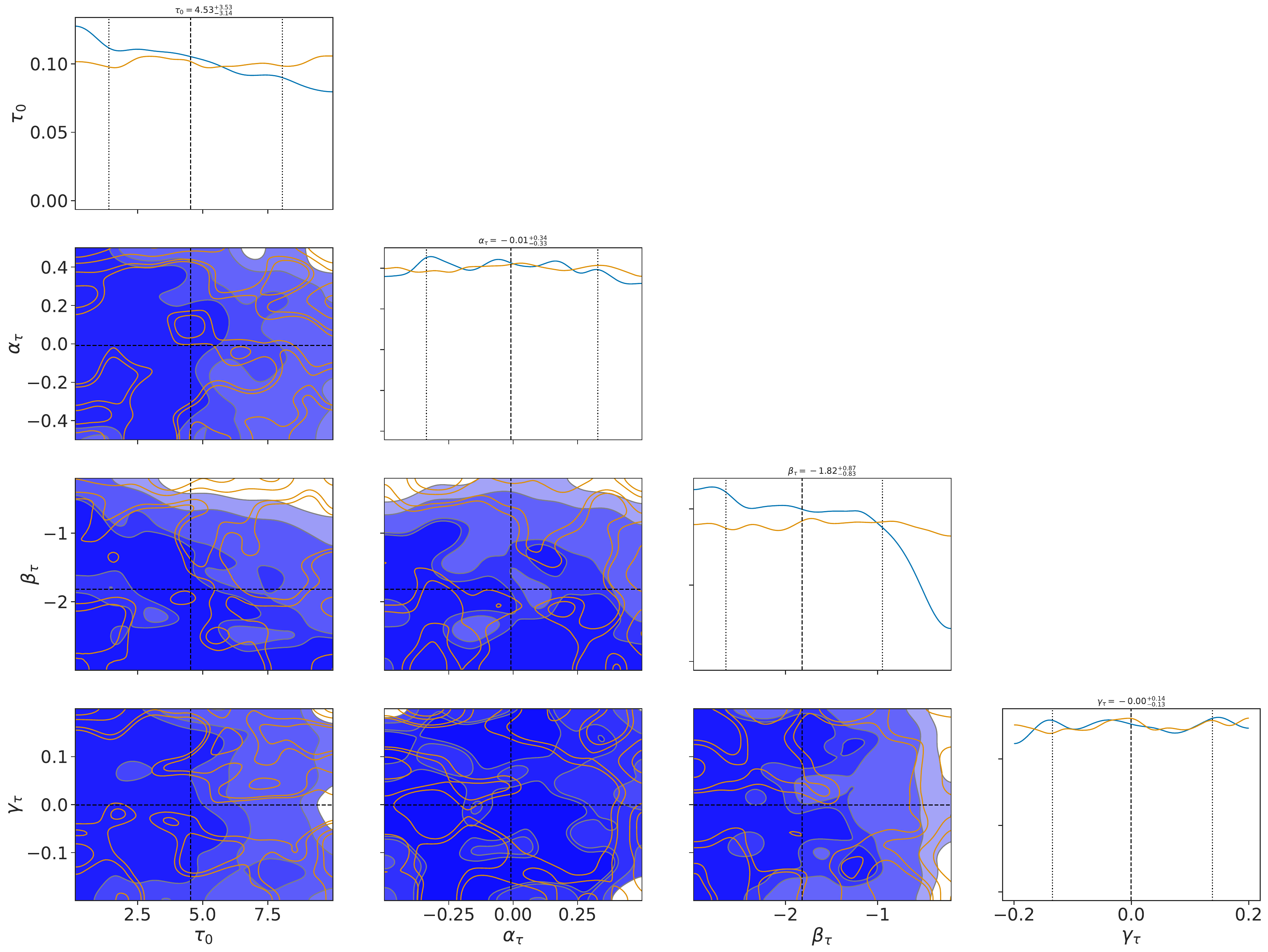}{0.45\textwidth}{(d) Merger timescale parameters.}}
    \caption{Close-ups of the prior (orange) and posterior (blue) distributions of parameters describing astrophysical observables, i.e. the galaxy stellar mass function (a), the $M_{\mathrm{BH}} - M_{\mathrm{bulge}}$ relation (b), $\mathcal{F}_{\mathrm{G, m}}$ (c), and $\tau_{\mathrm{m}}$ (d). Vertical dashed black lines show the median of each parameter, while vertical dotted lines show $68 \%$ confidence intervals.}
    \label{fig:observables}
\end{figure*}

We use a recent measurement of the GWB \citepalias{agazie_nanograv_2023} to constrain $\dot{\phi}_{\mathrm{BHB}}$ using MCMC.
We model $\dot{\phi}_{\mathrm{BHB}}$ as in \citetalias{chen_constraining_2019}, which assumes a galaxy stellar mass function (\autoref{eq:gsmf}), a galaxy pair fraction, and a galaxy merger timescale.
We model the fractional galaxy pairing rate as
\begin{equation}
\label{eq:gal_frac}
\dot{\mathcal{F}}_{\mathrm{p}}(M_{*}, z_{\mathrm{p}}, q_{*}) = \dot{f}_{0} \left(\frac{M_{*}}{a M_{0}} \right)^{\alpha_{f}} (1 + z_{\mathrm{p}})^{\beta_{f}} q^{\gamma_{f}} \, ,
\end{equation}
where $\dot{f}_{0}$ is the local fractional pairing rate of galaxies in major mergers at arbitrary reference mass $a M_{0} = 10^{11} \; \mathrm{M}_{\odot}$, and $\alpha_{f}$, $\beta_{f}$, and $\gamma_{f}$ determine how $\mathcal{F}_{\mathrm{p}}$ varies with $M_{*}$, $z_{\mathrm{p}}$, and $q_{*}$, respectively.
We similarly model the galaxy merger timescale as
\begin{equation}
\tau_{\mathrm{m}}(M_{*}, z_{\mathrm{p}}, q_{*}) = \tau_{0} \left(\frac{M_{*}}{b M_{0}} \right)^{\alpha_{\tau}} (1 + z_{\mathrm{p}})^{\beta_{\tau}} q^{\gamma_{\tau}} \, ,
\end{equation}
where $b M_{0} = 4 / h_{0} \times 10^{10} \; \mathrm{M}_{\odot}$, and $\tau_{0}$, $\alpha_{\tau}$, $\beta_{\tau}$, and $\gamma_{\tau}$ are defined analogously to their counterparts in \autoref{eq:gal_frac}.

For our fit we use astrophysical priors comparable to those used in previous studies (\citetalias{chen_constraining_2019}, \citealp{middleton_massive_2021}, \citealp{bi_implications_2023}).
These are derived from compiled observations of the galaxy stellar mass function (\citealp{perez-gonzalez_exploring_2008,muzzin_evolution_2013,tomczak_galaxy_2014,fontana_galaxy_2006,pozzetti_vimos_2007,kajisawa_moircs_2009,mortlock_deep_2011}, compiled in \citealp{conselice_evolution_2016}) and the $M_{\rm BH} - M_{\rm bulge}$ relation (\citealp{haring_black_2004,sani_spitzer_2011,beifiori_correlations_2012,mcconnell_revisiting_2013,graham_breaking_2012,kormendy_coevolution_2013}, compiled in \citealp{middleton_no_2018}).
% We assume uniform priors for $\mathcal{F}_{\mathrm{G, p}}$ and $\tau_{\mathrm{m}}$.
We assume uniform priors for the galaxy pair fraction and merger timescale.

We derive prior distributions for the galaxy stellar mass function from those compiled by \citet{conselice_evolution_2016}.
To do so we first fit \autoref{eq:gsmf} to the tabulated results from each of the compiled studies.
We next sample the posterior of each of these fits, combining all samples into a single dataset.
From this dataset we then estimate a multivariate normal distribution which we take as the galaxy stellar mass function prior for our fit of $\dot{\phi}_{\mathrm{BHB}}$.
We follow a similar procedure for our $M_{\mathrm{BH}} - M_{\mathrm{bulge}}$ priors, though we are able to skip the initial step of fitting tabular data to a model.

\begin{figure*}[!htb]
        \centering
        \includegraphics[width=0.8\linewidth]{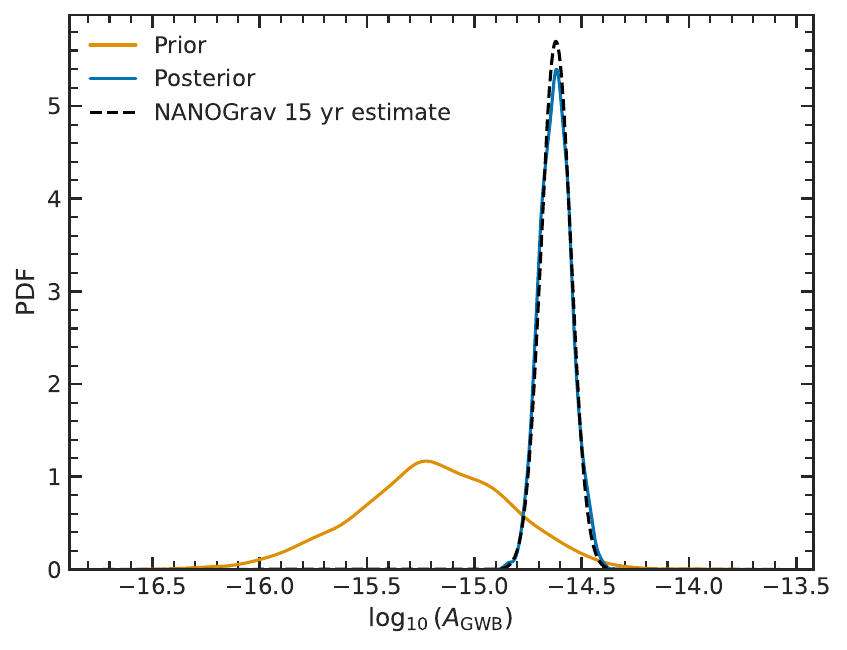}
        \caption{Prior (orange) and posterior (blue) distributions of the GWB amplitude predicted by our model. The target normal distribution of our fit (black) is derived from the GWB amplitude estimate in \citetalias{agazie_nanograv_2023}. The median and $90 \%$ confidence interval of the prior distribution is $A_{\mathrm{GWB}} = \left(6.8^{+17.2}_{-5.4}\right) \times 10^{-16}$, while for the posterior distribution we find $A_{\mathrm{GWB}} = \left(2.4^{+0.7}_{-0.6}\right) \times 10^{-15}$, consistent with \citetalias{agazie_nanograv_2023}.}
        \label{fig:gwb_comparison}
\end{figure*}

We then fit $\dot{\phi}_{\mathrm{BHB}}$ using these astrophysically motivated priors on the galaxy stellar mass function and $M_{\mathrm{BH}} - M_{\mathrm{bulge}}$ and uniform priors otherwise.
The posteriors of parameters describing specific astrophysical observables (i.e., the galaxy stellar mass function, the $M_{\mathrm{BH}} - M_{\mathrm{bulge}}$ relation, $\mathcal{F}_{\mathrm{G, p}}$, and $\tau_{\mathrm{m}}$) are shown in \autoref{fig:observables}.
We also show the prior- and posterior-predictive GWB amplitude distributions in Figure~\ref{fig:gwb_comparison}.
Our posterior distributions are comparable to those presented in previous studies (\citetalias{chen_constraining_2019}, \citealp{middleton_massive_2021}, \citealp{bi_implications_2023}).
We find $\dot{f}_{0} = 0.04 \pm 0.01$.
For comparison, \citep{casteels_galaxy_2014} use the Galaxy and Mass Assembly (GAMA) survey to estimate $\dot{f}_{0} \approx 0.016 - 0.046 \; \mathrm{Gyr}^{-1}$ for galaxies with $10^{9.5} < M_{*} < 10^{11.5} \; \mathrm{M}_{\odot}$, consistent with our results.
Similarly, \citet{rodriguez-gomez_merger_2015} use Illustris to estimate $\dot{f}_{0} \approx 0.02 - 0.03 \; \mathrm{Gyr}^{-1}$ for galaxies with $M_{*} \geq 10^{10} \; \mathrm{M}_{\odot}$, also consistent with our estimate.

To check the convergence of our MCMC chains we use a Gelman-Rubin statistic, which is a ratio comparing the variance of samples within each individual MCMC chain to the variance between chains \citep{gelman_inference_1992,vehtari_ranknormalization_2021}.
For converged MCMC chains, we expect the Gelman-Rubin statistic for each parameter to be less than 1.1, or less than 1.01 for a more conservative check \citep{vats_revisiting_2018}.
For our chains, all parameters have Gelman-Rubin statistics $\ll 1.01$, with the largest statistic being 1.002. We thus consider our MCMC chains converged.
Additional independent MCMC fits give similar results, so we consider the parameter space to be fully explored.

\pagebreak
\bibliography{bib}
\bibliographystyle{aasjournal}

\end{document}